\documentclass{jpconf}
\usepackage{graphicx}
\usepackage{amsmath}
\usepackage{amssymb}
\usepackage{lmodern}

\newcommand{\C}{\mathcal{C}}
\newcommand{\D}{\mathcal{D}}

\begin{document}
\title{On the geometric structures in evolutionary games on square and triangular lattices}

\author{Evgeni Burovski$^{1, 2}$, Aleksandr Malyutin$^{1,2}$, Lev Shchur$^{1, 2}$}
\address{$^1$ National Research University Higher School of Economics, 101000 Moscow, Russia}
\address{$^2$ Science Center in Chernogolovka, 142432 Chernogolovka, Russia} 

\begin{abstract}
We study a model of a spatial evolutionary game, based on the Prisoner's dilemma for two regular arrangements of players, on a square lattice and on a triangular lattice. We analyze steady state distributions of players which evolve from irregular, random initial configurations. We find significant differences between the square and triangular lattice, and we characterize the geometric structures which emerge on the triangular lattice.
\end{abstract}

\section{Introduction} 

Methods of statistical physics, originally developed for describing collective
phenomena in macroscopic collections of atoms and molecules, find their uses in studying
a broad range of phenomena, which, traditionally, would be considered out of scope. 
Some problems with origins in social sciences can be cast in the language of the
emergence of collective behavior from elementary interactions between individual agents. 
\cite{Perc2017}
A prominent one among these is the so-called evolution of cooperation \cite{Axelrod06}, i.e., the emergence of cooperative behavior where individual interactions favor competition.

A prototypical model where cooperation emerges from pairwise interactions is
provided by the classic Prisoner's dilemma game. Its spatially extended versions,
where a macroscopic number of agents repeatedly interact via the rules of the Prisoner's dilemma contain surprisingly rich variety of behaviors, including scenarios where
cooperation evolves in the long-time limit \cite{Smith82, Weibull95}.

A very simple yet highly non-trivial model, brought forward by Nowak and May \cite{Nowak1992, Nowak2006}, defines the synchronous, discrete-time, deterministic evolution of an ensemble of agents, which are arranged in a regular pattern. Despite its apparent
simplicity, the model displays surprisingly rich behavior and contains a series
of dynamic regimes, separated by sharp transitions. 

A natural question to ask is what is the role of the geometry of the arrangement of players. Original work \cite{Nowak1992, Nowak1993}, and recent efforts \cite{Kolotev2018, Kolotev_CSP2017}
mostly concentrated on the square grid geometry, where agents are located at the
sites of a square lattice in two spatial dimensions. \emph{A priori}, it is not
immediately clear whether a different local connectivity has a qualitative effect: on one hand large-scale structures should be insensitive to the local details; on the other
hand there are examples (e.g., variants of percolation problem, \cite{Adler})
where processes on square and triangular lattice display significant differences.

In this paper, we thus compare the dynamics of the evolutionary game on the square
and triangular lattices. The paper is organized as follows. In Section~\ref{sec:model} we describe the model and review some previously known results. Section~\ref{sec:steady-state} contains results of simulations and data analysis. In Section~\ref{sec:conclusion} we discuss results and future work.

\section{Model of spatial evolution game}
\label{sec:model}

A prototypical model of the game theory is the so-called \emph{Prisoner's dilemma} (PD),
played by two players in discrete time steps. In each round of the game, a
player uses one of two possible strategies, \emph{cooperate}, $\mathcal{C}$, or
\emph{defect}, $\mathcal{D}$, and receives a payoff which depends
on the strategies of the player and its opponent~\cite{Tadelis2013}.  We use the following payoff structure
\cite{Nowak1992, Nowak1993}: (i) If two players are $\mathcal{D}$ , they receive nothing; 
(ii) If both players $\mathcal{C}$, each of them receives a payoff of $S$, which
we set to $S=1$ without loss of generality; (iii) In the interaction of
$\mathcal{C}$ and $\mathcal{D}$, the $\mathcal{D}$ receives a payoff $T > S$ and the
$\mathcal{C}$ receives zero. This way, the payoff structure only depends on a
single parameter, the payoff parameter $b = T/S$.

For the spatial version of the game \cite{Nowak1992, Nowak1993}, we consider a
collection of players arranged in the vertices of a regular grid. In each round, 
an agent plays pairwise games with its neighbors, and the total payoff of an agent
is the sum of the pair-wise payoffs. The game is globally synchronous, so that
the strategy of a player is only updated after all pairwise games are finished
and all payoffs for all players are known.

Once all payoffs are known, players update their strategies: each agent adopts
the strategy which led to the maximum payoff \emph{among its neighbors}. This
concludes the time step, and the game repeats with the new distribution of strategies.

We consider two arrangements of players. First, we consider a square grid,
where players are arranged in the sites of a square lattice.
In a round, an agent plays nine games: with itself and its eight neighbors 
(chess king's moves).  We include self-interaction for
consistency with earlier work  \cite{Nowak1992, Nowak1993}.
Second, we consider the triangular lattice arrangement. Since the triangular lattice
can be thought of as a square lattice with additional diagonal bonds
(SW and NE, but not SE and NW), a round of the game has an agent interacting
with its six neighbors (and itself).

The evolution of the game can be thought of as a specific kind of a cellular
automation \cite{Shiff}. We note however that in the language of cellular automata, these
games have rather large transition tables: the state of a player at time step
$t$ depends on the states of its next-to-next-nearest neighbors at time step $t-1$.
This way, the transition table for the square grid game has $2^{25}$ rules,
and $2^{19}$ rules for the triangular grid.

The dynamics of the game is governed by the payoff ratio, $b$. The discrete structure
of payoffs gives rise to a series of dynamic regimes separated by transitions
at discrete values of $b$. Given an initial configuration, the dynamics of the
game is exactly identical for all values of $b$ between two consecutive transitions.

The transition points themselves are simply rational fractions with the numerator
and denominator between one and the number of pairwise games played by an agent
(i.e., the number of nearest neighbors on the lattice plus one for self-interaction.)
For the square lattice, the candidate fractions between $1 < b < 3$ are
$$
9/8, 8/7, 7/6, 6/5, 5/4, 9/7, 4/3, 7/5, 3/2, 8/5, 5/3, 7/4, 9/5, 2, 9/4, 7/3, 5/2, 8/3.
$$

On the triangular lattice, the transitions are located at
$$
7/6, 6/5, 5/4, 4/3, 7/5, 3/2, 5/3, 7/4, 2, 7/3, 5/2.
$$

\section{Dynamic regimes and steady states of the game}
\label{sec:steady-state}

\begin{figure}[!ht]
\includegraphics[width=0.5\columnwidth, keepaspectratio=True]{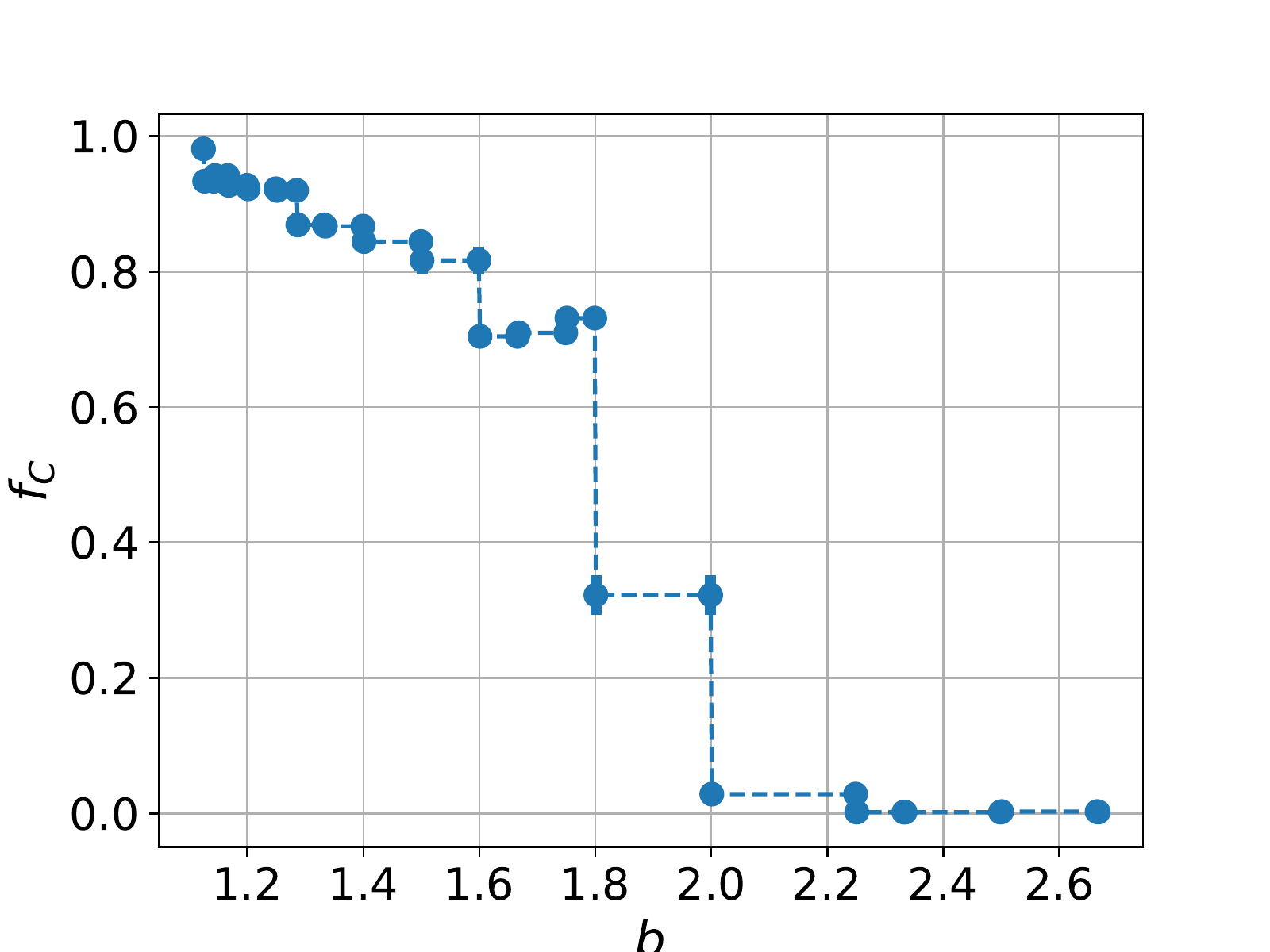}~\includegraphics[width=0.5\columnwidth, keepaspectratio=True]{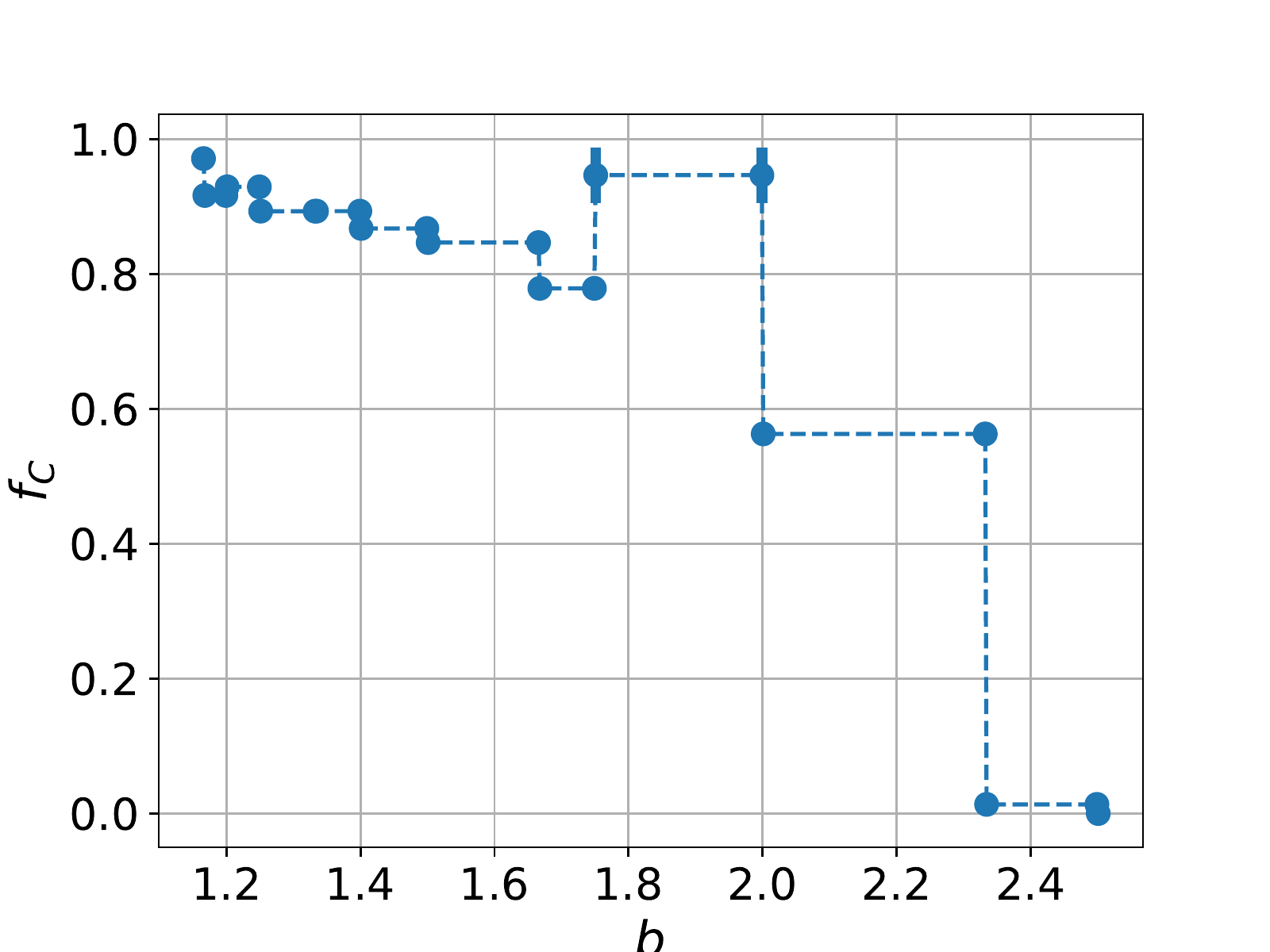}
\caption{Concentration of cooperators $\mathcal C$, $f_C$, as a function of the payoff parameter $b$. \textbf{Left}: square lattice. \textbf{Right}: triangular lattice. Each point is an average over
10 measurements taken every 1000 steps after 1000 burn-in steps for 10 independent realizations of the initial conditions with the initial density of $\C$, $p_0 = 0.1$. Dashed lines are to guide an eye. The game field size is $L=60$. See text for discussion.}
\label{fig:density}
\end{figure}

We consider a finite game field of the size $L\times L$ with periodic boundary
conditions in both directions for both square and triangular lattices. We start 
from random configuration, where at $t=0$ an agent is $\C$ with probability
$p_0 = 0.1$. We directly simulate the game numerically. We let the system evolve
for a certain number of burn-in steps (we typically take 1000 steps for the burn-in).
Fig.\ \ref{fig:density} shows the steady state average density of cooperators, $f_\C$,
(i.e., the number of cooperators divided by the total number of sites, $L^2$)
as a function of the payoff parameter $b$. Transitions are clearly manifest as
sharp jumps in the density of the components. To gain insight into the nature of
the transitions, it is instructive to consider small isolated objects, ``islands'' of one
strategy immersed into the sea of the other strategy.

\subsection{Square lattice}
\label{sec:square-lattice}

This geometry was discussed in detail in \cite{Nowak1993}, so here we only 
mention the key findings. For $b < 1$, cooperators always win. Defectors win
unconditionally for $b > 3$. A single isolated defector is stable for 
$1 < b < 9/8$.

For $b > 9/8$, a variety of lattice animals is possible: there are
various gliders, rotators, breathers and so on. Consider a straight $\D$
line one cell thick, separating two half-planes  of $\C$. It grows in the
perpendicular direction for $b > 3/2$, and becomes three
sites thick at the next step. A straight line interface between $\C$ and $\D$ grows into the
direction of $C$ for $b > 3$, while cooperators invade the $\D$ region
across the straight line interface for $b < 2$.
This way, for $3/2 < b < 2$, a straight line of $\D$ alternates between
thickness of one and three $\D$ with the period of 2 time steps.

A corner of $\D$ grows outwards into the $\C$ region for $b > 9/5$. Therefore, in the
regime of $9/5 < b < 2$, a square $3\times 3$ cluster of $\D$ grows from the corners
and shrinks in the middle of each face. This way, for $9/5 < b < 2$, an isolated
$\D$ generates a series of self-similar patterns, ``evolutionary rugs'', with
the mean density $f_C = 12\log{2} - 8$ \cite{Nowak1993}.

\subsection{Triangular lattice}
\label{sec:triangular-lattice}

\begin{figure}
\begin{center}
\includegraphics[width=0.38\textwidth, keepaspectratio=True]{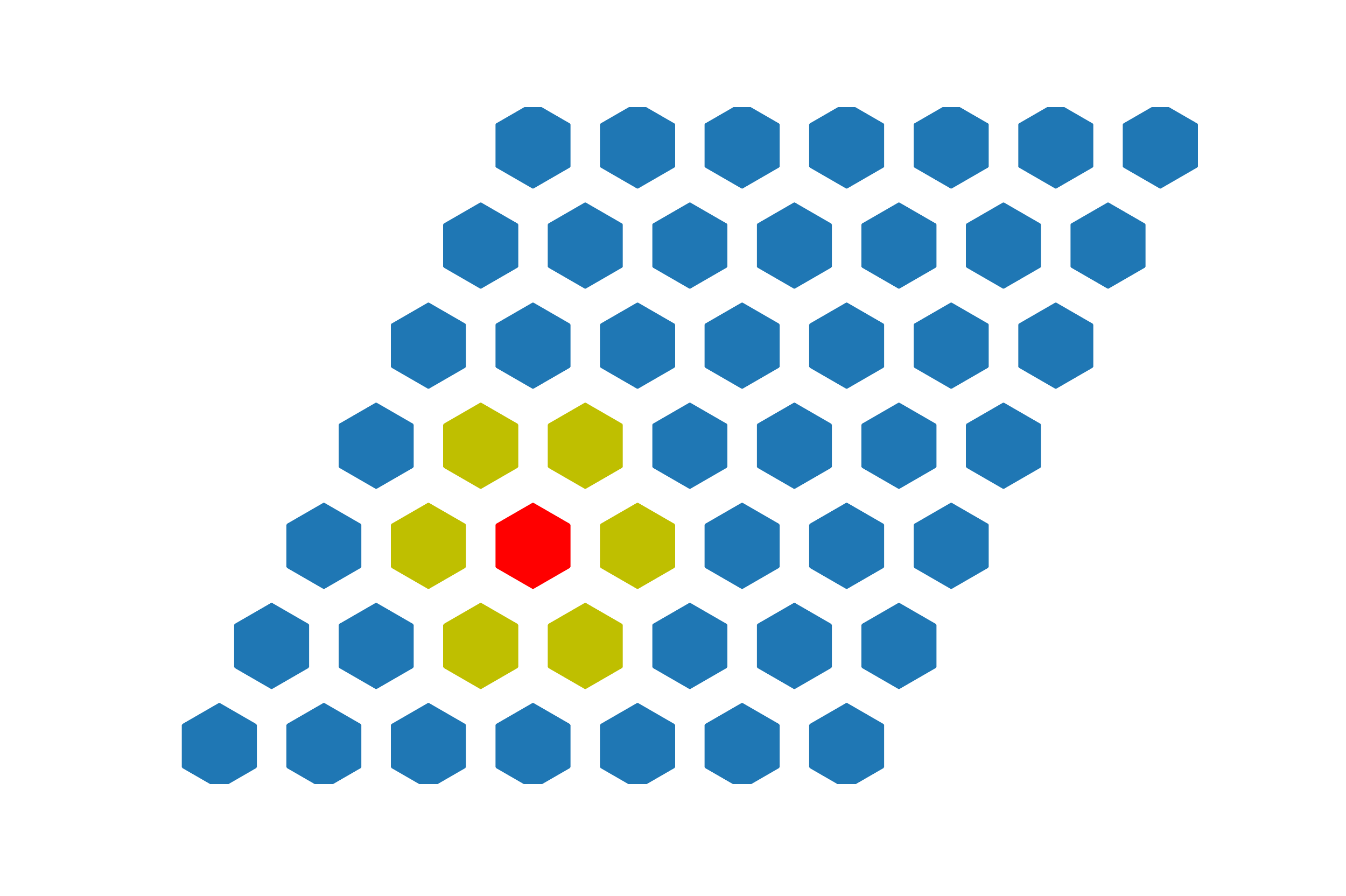}~%
\includegraphics[width=0.4\textwidth, keepaspectratio=True]{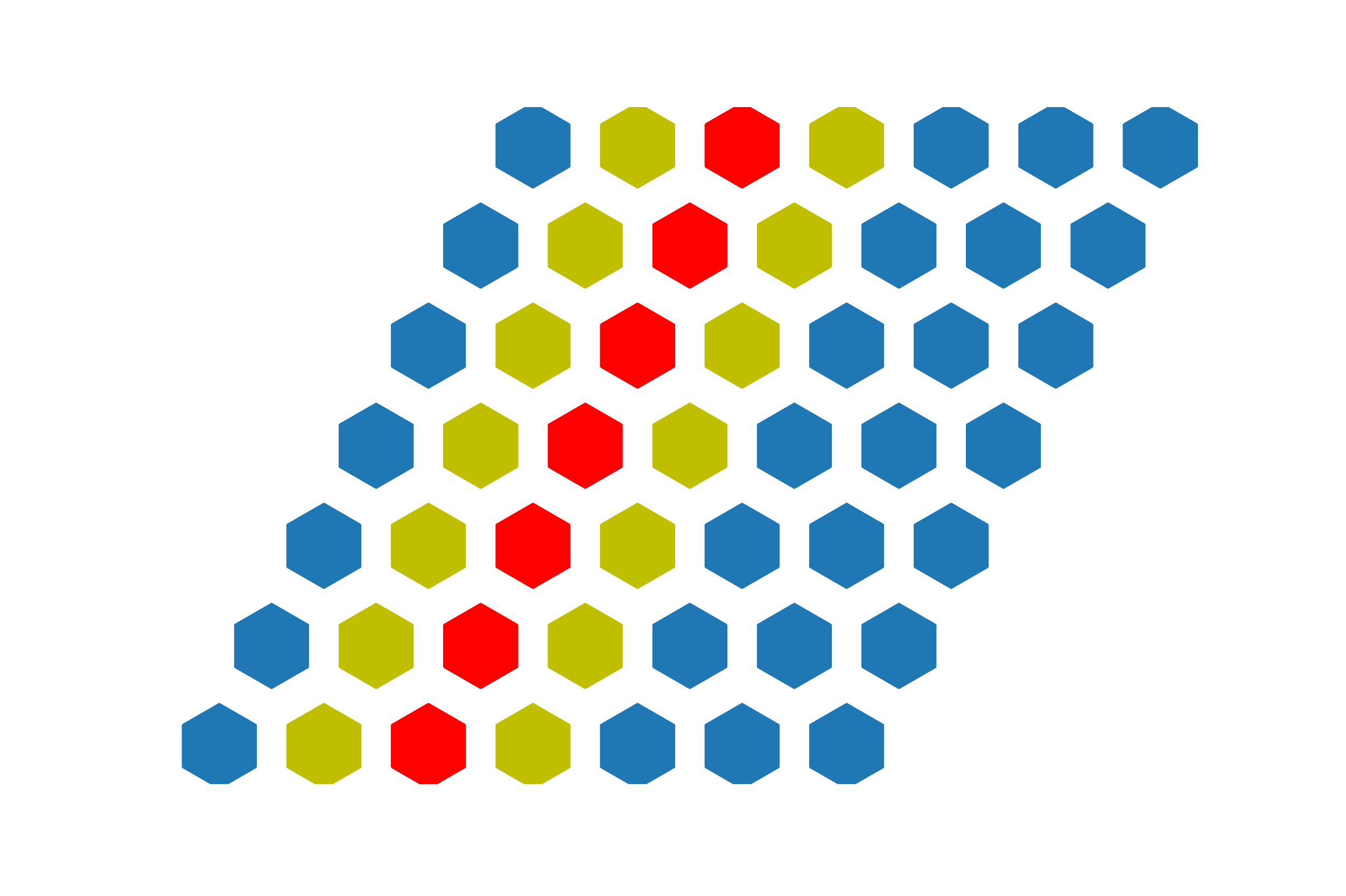}~%
\end{center}
\caption{(Color online.) \textbf{Left:} The cluster of seven $\D$. In a single time step, an
isolated $\D$ (red) grows to a cluster of seven sites (yellow) for $b > 7/6$.
The cluster becomes a blinker with period of one time step
for $7/6 < b < 2$, is stable for $2 < b < 7/3$ and grows without bound for $b > 7/3$.
\textbf{Right:} Time evolution of a straight line of $\D$. See text for discussion. 
Color coding is consistent with Fig.\ \ref{fig:snapshots}
}
\label{fig:seven_D}
\end{figure}

The analysis of specific isolated objects is similar to that on the square
lattice, Sec.\ \ref{sec:square-lattice}.
An isolated defector disappears after a single step for $b < 1$. For $1 < b < 7/6$
a single defector is stable. For $b > 7/6$ a single defector grows to a cluster of
seven $\D$, see Fig.\ \ref{fig:seven_D}. Further evolution of the seven-$\D$ cluster
depends on $b$: for $b < 6/3$, the cluster shrinks back to a single $\D$ at the center, 
so that we have a blinker of the size alternating between 1 and 7. For $b > 7/3$,
it keeps growing, while for $2 < b < 7/3$ a seven-$\D$ cluster is stable.

A straight line of $\D$ of unit thickness disappears if $b < 5/4$. Otherwise, 
it remains stable for $5/4 < b < 7/4$, and grows for $b > 7/4$,
see Fig.\ \ref{fig:seven_D}. The result of the latter case, a line of $\D$
of thickness three, remains stable for $ 5/2 < b < 7/2$, and shrinks back for the values
of $b$ below that range. Therefore, a straight line blinker with unit period 
exists in the range $7/4 < b < 5/2$.

\subsection{Unstructured initial conditions: square lattice}
\label{sec:snapshots-square}

\begin{figure*}[!b]
\center
\includegraphics[width=0.25\textwidth, keepaspectratio=True]{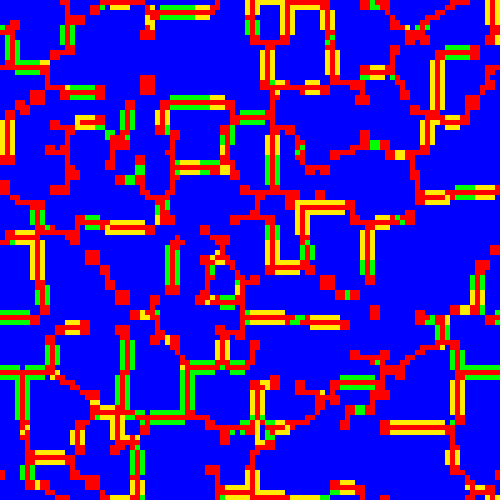}~~~%
\includegraphics[width=0.25\textwidth, keepaspectratio=True]{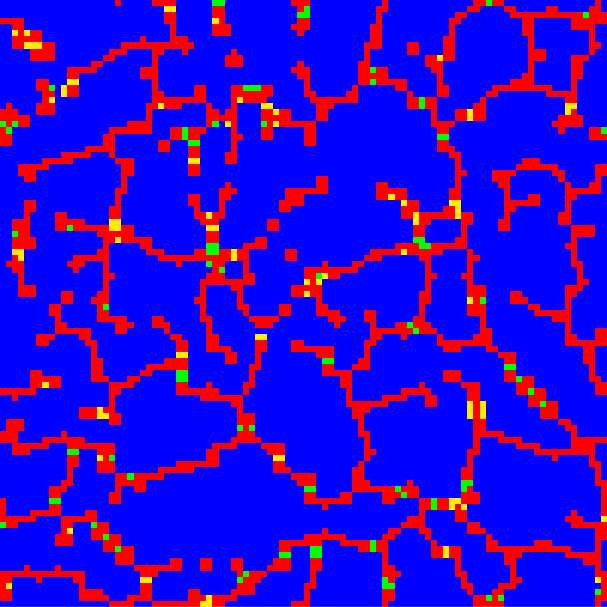}~~~%
\includegraphics[width=0.25\textwidth, keepaspectratio=True]{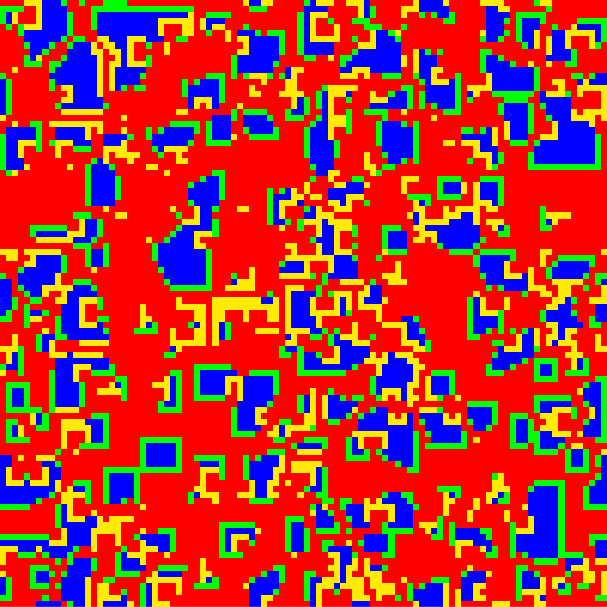}
\caption{(Color online.) Representative snapshots of the game field for
$b=1.74$ (left), $b=1.79$ (center) and $b=1.81$ (right). The color coding is
consistent with Ref.\ \cite{Nowak1992, Nowak1993}: blue is $\mathcal{C}$, red is
$\mathcal{D}$, yellow is a $\mathcal{D}$ that was
a $\mathcal{C}$ in the preceding round, and green is a $\mathcal{C}$ that was
a $\mathcal{D}$ in the preceding round \cite{Kolotev2018}. See text for discussion.
}
\label{fig:snapshots}
\end{figure*}

Fig.\ \ref{fig:snapshots} shows representative snapshots of the game field in the
steady state which develops from a random initial configuration with $p_0 = 0.1$
for several values of the payoff parameter $b$. For $b < 9/5$, the system develops
relatively static web-like structures of defectors. The most non-trivial regime
is $9/5 < b < 2$. In this regime both kinds of
strategies grow into regions of the opposite strategy, leading to the chaotic
evolution where the state of the game changes significantly at the time scale
of several time steps, and the mean density of cooperators fluctuates around the
$f_C \approx 0.31$ (see Sec \ref{sec:square-lattice}). \cite{Nowak1992}.

\subsection{Unstructured initial conditions: triangular lattice}
\label{sec:snapshots-triangular}

\begin{figure*}
\center
\includegraphics[width=0.34\textwidth, keepaspectratio=True]{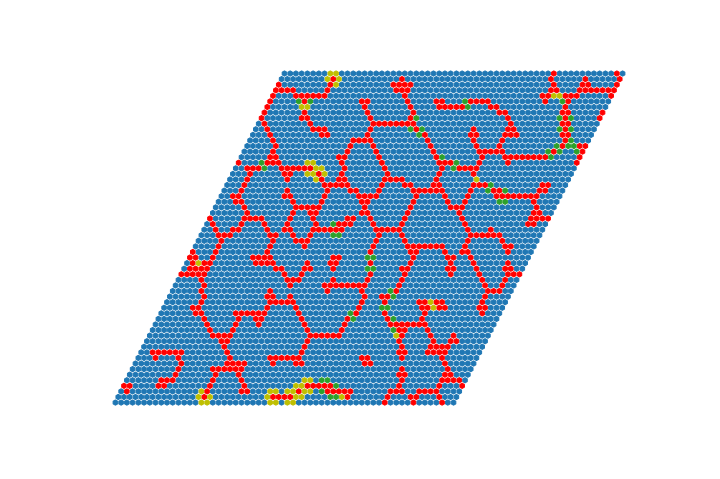}~%
\includegraphics[width=0.34\textwidth, keepaspectratio=True]{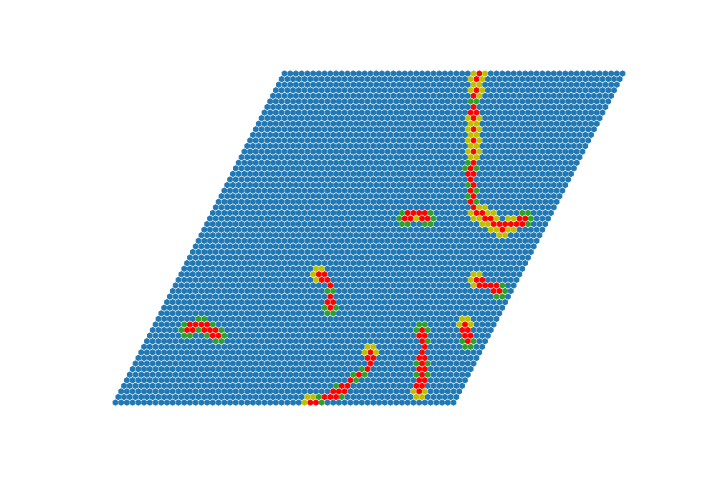}~%
\includegraphics[width=0.34\textwidth, keepaspectratio=True]{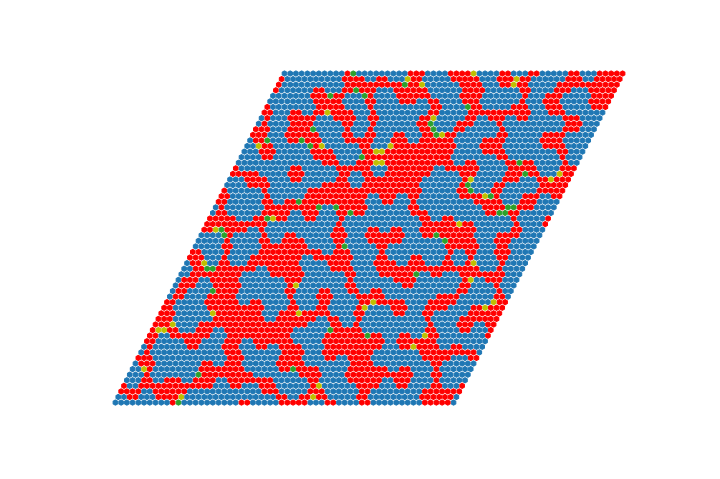}
\caption{(Color online.) Representative snapshots of the triangular grid game
field for $b \in (5/3, 7/4)$ (left), $b \in (7/4, 2)$ (center) and $b \in (2, 7/3)$ (right). Here the field size is $L=60$. The color coding is
consistent with Fig.\ \ref{fig:snapshots}. See text for discussion.
}
\label{fig:snapshots_triangular}
\end{figure*}

Fig.\ \ref{fig:snapshots_triangular} shows representative snapshots of the
game field in the steady state which develops from a random initial
configuration with $p_0 = 0.1$ for several values of the payoff parameter $b$. 
Typical steady state distributions of strategies strongly depend on the payoff
parameter.

For $2 < b < 7/3$, the system develops an unstructured pattern of
$\C$ and $\D$, which can be thought of as a random collection of seven-$\D$
clusters (cf Sec.\ \ref{sec:triangular-lattice}), arranged in random ways, which are dictated by a specific realization
of initial conditions. The pattern is mostly static, with small-scale oscillations
at the boundaries.

Across the transition at $b = 2$, typical arrangements of the game field are
very different. The field is dominated by cooperators, with relatively small
isolated regions of $\D$ blinkers consisting of a static core $\D$ region,
surrounded  by a layer of ``swingers'', which frequently change their strategy.

For $5/3 < b < 7/4$, typical patters are also different. In this regime,
defectors also arrange in static web-like ``backbone'' which is interconnected
and spans the system size. (This is in contrast to regime of $7/4 < b < 2$, where $\D$
clusters are isolated.) Relatively small clusters of ``swingers'', where strategies oscillate with periods of the order of several time steps attach to this static web of $\D$.

\section{Switching times}
\label{sec:switching-times}

To further quantify the dynamics of the game, we study the mean time for 
for an agent to change their strategy. Specifically, we perform the following
simulation: Starting from a random configuration, we evolve it for $10^4$ steps
(after $10^3$ burn-in steps). For each player we record the time steps where
they switched the strategy, and compute the mean time between switches. 
Results are summarized in Table \ref{tab:switches}. For each range of the 
payoff parameter $b$ we show the number of players who switched their strategy
at least once, $n_{sw}$, and the mean time between switches \emph{for these 
players}, $\tau_{sw}$.

For the square lattice, $b \in (9/5, 2)$ stands out: all $L^2$ players change
their strategies with the time scale of around four steps. All other regimes are
much more static, with a relatively small fraction of ``swinging players''
who change the strategy every time step on average.

On the triangular lattice, the special regime is $7/4 < b < 2$, where a sizable
fraction of players engage in relatively slow processes, with time scales much
larger then one. We note that the time scales are comparable to the simulation
times, therefore we cannot yet rule out that what we observe is some very slow
transient phenomenon. All other regimes are static, with only a small fraction of 
fast ``swinging'' players.

\begin{table}
\begin{center}
\begin{tabular}{ccc|ccc}
\multicolumn{3}{c}{Triangular lattice} & \multicolumn{3}{c}{Square lattice}\\[2pt]
$b$ range & $n_{sw}$ & $\tau_{sw}$ & $b$ range  & $n_{sw}$ & $\tau_{sw}$ \\
\hline
$5/4$ &  92    &  1.0        & 4/3  & 99       &  1.1 \\
$4/3$ &  12    &  1.0        & 7/5  & 7        &  1.5 \\
$7/5$ &  153   &  1.0        & 3/2  & 233      &  1.5 \\
$3/2$ &  64    &  1.1        & 8/5  & 148      &  1.1 \\
$5/3$ &  57    &  1.7        & 5/3  & 105      &  1.1 \\
$7/4$ &  348   &  590      & 7/4  & 13       &  1.0 \\
$2$   &  35    &  1.3        & 9/5   &  900    &  4.1 \\       
\end{tabular}
\end{center}
\caption{Switching times and numbers for the square and triangular lattice. 
The simulation is $10^4$ steps on lattices with $L=30$. $n_{sw}$ is the number
of players, which switched their strategy more then once, and $\tau_{sw}$ is the
mean time between switches for these $n_{sw}$ players. Each row shows results for
the payoff parameter between the value in the row and the next row (e.g.,
``$b$ range 5/4'' means $5/4 < b < 4/3$). See text for discussion.}
\label{tab:switches}
\end{table}

\section{Conclusions and outlook}
\label{sec:conclusion}

We study an evolutionary game based on the Prisoner's Dilemma with regular
arrangements of agents in a plane. Despite its apparent simplicity, this deterministic, globally synchronous game displays surprisingly rich behavior in the long time limit. We compare the steady states of the game, for two  ways of arranging the players, on the square grid and on the triangular grid.  The local structure of connections between neighboring players significantly affects the possible steady states: for the square grid,
there exists a chaotic regime; on the triangular lattice, steady states are mostly static. However, these static patterns are significantly different for different values of
the payoff parameter. 
It would be interesting further characterize these emergent geometric structures
and their oscillatory patterns of evolution, which we intend to do in a future study.

\ack
This work was partially supported by RFBR grant 16-07-01122 (development of the simulation algorithms). A.M. acknowledges the support of the Academic Fund Program at the National
Research University Higher School of Economics (HSE) in 2018-2019 (grant No 18-05-0024) and by the Russian Academic Excellence Project ``5-100''.


\end{document}